\documentclass{ws-procs975x65}
\usepackage{latexsym}
\usepackage{amssymb}
\usepackage{amsmath}
\usepackage{float}
\usepackage{indentfirst}
\usepackage{array}
\usepackage{xspace}
\usepackage{amscd}

\newcommand{\kn}{{\kern 1pt}}
\newcommand{\akn}{{\kern -1pt}}

\newcommand{\vm}[1]{\vphantom{#1}}
\newcommand{\hsp}{\hspace{0.15em}}
\newcommand{\nhsp}{\hspace{-0.15em}}
\newcommand{\smhsp}{\hspace{0.06em}}

\newcommand{\ds}{\displaystyle}

\newcommand{\vak}{\varkappa\,}
\newcommand{\pa}{{\partial}}
\newcommand{\ga}{\gamma}
\newcommand{\si}{\sigma}
\newcommand{\Si}{\Sigma}

\newcommand{\fr}{\frac}

\newcommand{\bw}{\begin{widetext}}
\newcommand{\ew}{\end{widetext}}
\newcommand{\be}{\begin{align}}
\newcommand{\ee}{\end{align}}
\newcommand{\ba}{\begin{eqnarray}}
\newcommand{\ea}{\end{eqnarray}}
\newcommand{\non}{\nonumber}

\newcommand{\h}{H}

\def\cV{{\cal V}}

\def\e{{\rm e}}

\newcommand{\sgn}{{\rm sgn\smhsp}}
\newcommand{\nn}{\nonumber}

\newcommand{\ah}{ \mathrm{a}}
\newcommand{\bh}{ \mathrm{b}}

\newcommand{\vp}{\vphantom{\frac{a}{a}}}

\begin{document}
\title{ DOMAIN WALLS: MOMENTUM CONSERVATION \\IN ABSENCE OF ASYMPTOTIC STATES}
\author{D.\,V.\,GAL'TSOV$^{1}$, E.\,Yu.\,MELKUMOVA$^{1}$, P.\,SPIRIN$^{1,2}$}
\address{\parbox{11cm}{\noindent\rule{0cm}{0.4cm}{}$^{1}$\,Faculty
of
Physics, Moscow State University, 119899, Moscow, Russia;\\
{}$^{2}$\,Institute of Theoretical and Computational Physics,
Department of Physics,\\ \phantom{$^{2}$}\,University of Crete, 70013 Heraklion,
Greece.} \\
\email{galtsov@phys.msu.ru, elenamelk@mail.ru,
pspirin@physics.uoc.gr}}
\begin{abstract}
Gravitational potentials of the domain walls in the linearized
gravity are growing with distance, so the particle scattering by the
wall can not be described in terms  of free asymptotic states. In
the non-relativistic case this problem is solved using the concept
of the potential energy. We show  that in the relativistic  case one
is able to introduce  gravitationally dressed   momenta the sum of
which is conserved up to the momentum flux through the lateral
surface of the world tube describing losses due to excitation of the
branon waves.
\end{abstract}
\bodymatter
\section{Introduction}\label{intro}

In the standard theory of particle collisions, both  classical and
quantum, one assumes the existence of asymptotic states in which the
particles can be regarded as non-interacting.  For this picture to
be valid, the interaction force between the colliding objects has to
fall down sufficiently fast with the distance.  Meanwhile, in
various physical systems, like quarks joined by the gluon strings,
this is not so,  and the question arises, whether one can sensibly
define the notion of the potential energy within the classical
relativistic two-body problem. With this motivation, we consider the
scattering problem for a point particle impinging onto the brane
imbedded into space-time with the codimension one, in which case the
interaction force does not fall down asymptotically. We will be
interested in the two-body problem, accounting for the brane
back-reaction on equal footing with the particle.
  This problem may have physical applications in
cosmology \cite{linear,Vilsh}{}, in particular, perforation of the
domain walls by black holes was suggested as a novel mechanism of
domain walls destruction in the Early Universe
\cite{ChamEard,Stojkovic:2005zh,Flachi:2007ev}{}. It may be of
interest also in the study of the black hole escape from the
Randall-Sundrum brane, in the dynamical description of colliding
branes in supergravity/string theory and so on.

 Recently we have
shown \cite{GaMeS1} that the  perforation of the domain wall by a
point particle can be described within linearized gravity in terms
of distributions.  Here we  concentrate on the  the energy-momentum
balance \cite{GaMeS2} in this process, extending the treatment to
the second order in gravitational constant. This allows to introduce
the effective gravitational stress-tensor which has to be taken into
account in the energy-momentum balance. This tensor generically is
non-local, but we show that it can be still unambiguously divided
between the two objects leading to the notion of gravitationally
dressed momenta whose balance involves an extra momentum flux
through the lateral surface of the world tube.

\section{The setup}
Our system consists of a point particle moving along the world-line
$x^M=z^M(\tau)$ and an infinitely thin domain wall filling the
world-volume $\cV_{D-1}$  given by the embedding equations
\mbox{ $x^M=X^M(\sigma^\mu)$ } in $D-$ dimensional space-time (\mbox{$D\geqslant 4$}) with the
metric $g_{MN}$, $M=0,...,D-1,\;\mu=0,...D-2$ of the signature
$(+,-,...,-)$. The action reads:
\begin{align}\label{Ac}
S= &-\frac{1}{2} \int  \left(e\; g_{MN}\dot{z}^M
\dot{z}^N+\frac{m^2}{e}\right)
d\tau\,\\
&-\fr{\mu}{2}\int\left[\vp
  X_\mu^M X_\nu^N g_{MN}\gamma^{\mu\nu}-(D-3)\right]\sqrt{-\gamma}\:d^{D-1}
  \sigma  -
  \frac{1}{\vak^2}\!\int\! R_D \,\sqrt{-g}\; d^D x& \non
\end{align}
where the first term is the particle action in the Polyakov form
($e(\tau)$ is the einbein on  the particle world-line), the second
one is the domain wall geometrical action, $X_\mu^M=\pa
X^M/\pa\hsp\sigma^\mu$ are the tangent vectors and
\mbox{ $\gamma^{\mu\nu}=X_\mu^M  X_\nu^N g_{MN}{\big
 |}_{x=X}$ } is the inverse metric on the domain wall
world-volume $\cV_{D-1}$, $\gamma={\rm det} \gamma_{\mu\nu}$. The
last term is the Einstein-Hilbert action, $ \vak^2\equiv 16\pi G_D
$.

 To treat the  problem perturbatively we expand all
variables in powers of $\vak$  and derive the system of iterative
equations. The $D-$dimensional cartesian coordinates of the
embedding space-time are split as $x^M=(x^{\mu},z)$,
$x^{\mu}=(t,\bf{r})$, and the particle is assumed to move along $z$,
i.e. normally to the domain wall. In the zeroth order the particle
is assumed to move with the constant velocity $u^M=\gamma(1, 0, ...,
0, v),$ \mbox{ $\gamma=1/\sqrt{1-v^2},$ } so the world-line is $ z^{M}(\tau)= u^M\tau$ while the
einbein  reads \mbox{ $e={\rm const}=m\,$, }
corresponding to the parametrization in terms of the proper time.
The wall in the zeroth order is assumed to be plane, unexcited and
being at rest at $z=0$ in the chosen Lorentz frame: $ X^M =\Si^M_\mu
\si^\mu\,, $
 where $\Si^M_\mu=\delta^M_\mu$ are $(D-1)$ constant Minkowski vectors
 normalized as $
\ga_{\mu\nu}=\eta_{\mu\nu}\,.$ The moment of perforation of the wall
by the particle that occurs at $z=0$ is $t=\tau=0$.

 The metric deviation must be further expanded in $\vak$:
 $$ \h^{MN}=h^{MN}+{\bar h}^{MN}+\delta  \h^{MN}\,,
 $$  where the
first order term is split into the sum of contributions of the wall
 \begin{align} \label{brgr}
h_{MN}= \frac{2 k \hsp |z|}{\vak}
\:{\rm diag}\hsp(-1,1,...,1,{D-1})\,, \qquad k \equiv \frac{\vak^2 \mu }{4(D-2)}
 \end{align}
  and of the particle
 \begin{align} \label{hpart}
 \bar{h}_{MN}=-\fr{\vak
\,m\Gamma\left(\fr{D-3}{2}\right)}{4\pi^{\fr{D-1}{2}}}
 \left(u_M
u_N-\fr{1}{D-2}\,\eta_{MN}\right)\fr{1}{[\gamma^2(z-v
t)^2+r^2]^{\fr{D-3}{2}}}\,.
 \end{align}
  The second order metric
deviation $\delta \h^{MN}$
 does not split  anymore on separate contributions and obeys (in
the same gauge) the d'Alembert equation
\begin{align}
  \Box \Bigl( \delta\h^{MN}-\frac{1}{2}\,\eta^{MN}\delta\h\Bigr) =  -\vak \left(\delta T^{MN}+ \delta{\bar T}^{MN}+S^{MN}(h,\bar
h)\right),
  \label{psi2eq}
\end{align}
where the perturbations of the  particle (noted with the bar)  and
the brane stress-tensors:
\begin{align}
\label{T1MN_r}  \delta \bar {T}^{MN}(x)= \frac{m}{2} \int \left[\hsp
4 \hsp  \delta\dot z^{(M} u^{N)} -\vak  \, u^M u^N \left( {h}+ 2
\hsp  \delta{z}^P\pa_P \vp \right) \right] \delta^D\!\left(x- u
\tau\right)\, d\tau\,,
\end{align}
with $h $ being the trace of the first order metric deviation due to
the wall (\ref{brgr}); the symmetrization $(MN)$  over the indices
is defined  with 1/2. The delta-function indicates on the
localization of the integrand at the non-perturbed particle
world-line.
\begin{align} \label{taumn}
  & \delta T^{ {M} {N}}  (x) =
\fr{\mu}{2} \int  \left[ 4\, \delta_{\mu}^{(M}\delta{\nhsp\smhsp
X}_{\vm{\mu} \nu}^{N)} \eta^{\mu\nu}   -2\, \delta^M_\mu
\delta^N_{\nu\vm{\mu}} \left( \bar{h}^{ \mu\nu}  + 2\, \eta^{LR}
\delta_R^{( \mu } \delta{\nhsp\smhsp X}_L^{\nu)} \right) + \right.
\nn\\ & \left. \;\;+
 \delta^M_{\mu}\delta^{N}_{\nu\vm{\mu}} \eta^{ \mu\nu} \left(\bar{h}^{\lambda}_{\lambda}-
\bar{h}+2\hsp \delta{\nhsp\smhsp X}^{L}_{\lambda}\delta^{\lambda}_L
-2\hsp \delta{\nhsp\smhsp X}^{L}
\partial_L \vp\right) \right]\delta^{D-1}\!\left(x -  \sigma \right)\,\delta(z)\: d^{D-1}\si
\,,
 \end{align}
 Again, the delta-functions in the integrand indicate on  its
localization on the unperturbed wall  world-volume. Finally
$S^{MN}(h,\bar h)$ stands for the quadratic form
\begin{align}
\label{natag_0} {{S}}^{MN} & =   2 \hsp  {\h}^{MP , Q}
\h^{N}{}_{\![Q , P]} + {\h}_{PQ}\left(\h^{MP ,NQ}+ \h^{NP , MQ}-
\h^{PQ, MN}- \h^{MN , PQ}\vp \right) \nn\\& -2\hsp \h^{(M}_{P}
\Box \h^{N)P}_{\phantom{P}} -
   -\frac{1}{2}\,  {\h}^{PQ  , M} \h_{PQ}{}^{,
N}+\frac{1}{2}\, {\h}^{MN}\Box \h  \nn\\ &+ \frac{1}{2}\,\eta^{MN}\Bigl(2
{\h}^{PQ}\Box \h_{PQ}- {\h}_{PQ , L} \h^{PL , Q}+\frac{3}{2}\,
{\h}_{PQ , L} \h^{PQ,L}\Bigr),
\end{align}
in which $\h^{MN}$ should be taken as the sum $\h^{MN}=h^{MN}+{\bar
h}^{MN}$, keeping only the crossed terms in $h^{MN}\,,{\bar
h}^{MN}$.

  Perturbation  of the domain wall embedding functions
$\delta X^M$ due to gravitational interaction with the particle  in
the aligned coordinates on the brane $\sigma^{\mu}=(t,\mathbf{r}) $
is described by a single component $\Phi=\delta X^z$ as follows
\cite{GaMeS1}:
\begin{align}\label{bran}
  \Box_{D-1} \Phi(\si^{\mu})=\vak
\left(\frac{1}{2}\,
 \eta_{{\mu\nu}}\bar{h}^{\hsp \mu\nu,z}-\bar{h}^{\hsp  z\hsp  0,0}\right)_{\!z=0}
,
 \end{align}
where $\Box_{D-1} \equiv \partial_{\mu} \partial^{\mu}$. The
retarded solution of this equation consists of two parts $\Phi=
\Phi_{\ah}+\Phi_{\smhsp\bh}$, where the first is antisymmetric   in
time and represents an eventual deformation of the wall correlated
with the particle motion.   The second part is the spherical branon
wave starting at the moment of perforation $
(\Phi_{\smhsp\bh}\sim\theta\kn(t)) $ and propagating to infinity
with the velocity of light \cite{GaMeS1}{}:
 \begin{align}
 &\Phi_{\ah}\equiv - \Lambda \,\sgn (t)\,I_{\ah}\,,\qquad \Phi_{\smhsp\bh}
\equiv 2\kn \Lambda  \, \theta(t)\,I_{\bh}\,, \qquad \Lambda = \fr{\vak^2
m \gamma^{-1} }{4(2\pi)^{D/2-1} }\Bigl(
\gamma^2v^2 +\fr{1}{D-2}\Bigr) \,,\nn\\
  & I_{\ah,\bh}(t,r) = \frac{1}{r^{\fr{D-4}{2}}} \int\limits_{0}^{\infty} \! dk
\,J_{\frac{D-4}{2}}(k r)\,{k}^{\frac{D-6}{2}} \,w_{\ah,\bh}(t,k)\,, \qquad
 w_{\ah,\bh}= \left\{ \begin{array}{c}
\e^{- k\gamma v|t|} \\  \cos \smhsp kt
\end{array} \right\}
\,, \nn
 \end{align}
where $J_{\nu}{(z)}$ is a Bessel function of the first kind.

\section{Conservation of the energy-momentum}
In the first order in $\vak$ the total energy-momentum tensor
consists of three contributions (\ref{psi2eq}) and satisfies the
conservation equation $ \partial_N\tau^{MN}=0 $. To convert the
latter into the energy-momentum balance equation one has to
integrate over the world-tube $\Omega$: $$
0=\int_\Omega\partial_N\tau^{MN}=\int_{\partial \Omega}\tau^{MN}d
\Sigma_N\,,
 $$ bounded by the closed hypersurface,
consisting of two space-like hypersurfaces associated with the
moments of time $t_0,\, t_f$ (usually chosen orthogonal to the
time-axis), and the closing lateral hypersurface $\Sigma_{\infty}$
at spatial infinity.

Since the wall is  infinite its total energy-momentum diverges both
in zero and the first order in $\vak$, so some subtraction is
required.   Another problem is the non-zero lateral flux through
$\Sigma_{\infty}$ from the wall. In accord with the  splitting of
the total energy-momentum tensor (\ref{psi2eq}) one can write
\begin{align}\label{mom}
 \quad P_{\rm tot}^M(t)=\delta\bar{P}^M(t)+\delta P^M(t)+S^M(t)\,,
 \end{align}
where
 \begin{align}\label{momenp}
\delta\bar{P}^M(t)=
 \int \delta\bar{T}^{M0}\,  dz\, d^{D-2}\mathbf{r}\,,
 \qquad \delta P^M(t)=  \int  \delta T^{M0}\,  dz\, d^{D-2}\mathbf{r}
  \end{align}
   are the first-order
kinetic momenta carried by the particle
 and the wall, and
 \begin{align}\label{momenpS}
  S^M(t)=  \int  \delta S^{M0}  \,  dz\, d^{D-2} \mathbf{r}\,\end{align}
 is the momentum carried by
the gravitational field. The lateral flux  can be split into similar
three contributions, revealing that only the wall $\delta T^{Mr}$
 does not vanish.
  To  get rid of infinities,  we pass to the time derivatives
of the partial momenta which are all finite:
\begin{align}\label{difcons}
\!\!\!\! \frac{d}{dt}\left(\delta\bar{P}^M(t)+\delta
P^M(t)+S^M(t)\right)= -\!\lim_{R\to
\infty}\biggl(\int_{-\infty}^{\infty}
dz\int_{S^{D-3}_{R}}\!\!\!\!\delta
T^{Mr}R^{D-3}d^{D-3}\Omega\biggr),
\end{align}
 where the integral at the right hand side represents the lateral momentum flux. This term looks like an external
force acting upon the system.

The first-order   particle stress tensor is determined  by the
world-line perturbation and the metric deviation due to the wall.
 Integrating over the spatial volume, one obtains the derivative of
the particle momentum $  \delta\dot{\bar{P}}^{M} $ (only $z$ and
$0$-components are non-zero):
 \begin{align}
 \delta\dot{\bar{P}}^{z}  = m k
 \left[(3D-2)\ga v^2+\ga^{-1}\vp \right] \sgn(t),\qquad
  \delta\dot{\bar{P}}^{0}  =2D m k  \ga v \,\sgn(t).\nn
 \end{align}
The first order stress-tensor of the wall is obtained by using the
first-order metric deviation due to the particle and by the
first-order perturbations of the wall world-volume. The
corresponding  derivatives read:
\begin{align}\label{kbterm0}
& \delta\dot{P}^0=- \gamma v \left(\vp(D-2)\gamma^2 v^2 + 2D-7
\right) m k
 \,\sgn(t),\\
  \label{kbtermz}
& \delta\dot{P}^z_{\ah}=-  mk\left(\vp (D-2)\gamma^2 v^2 +1
\right)\gamma v^2 \,\sgn(t)
    \,
 \non\\
&   \delta \dot{P}^z_{\bh}= -
  \fr{2 \hsp k m  }{\gamma }\left( (D-2)\,\gamma^2v^2+
1\vp \right)  \theta(t) \,.
\end{align} where $z$-components are split into $a\,,b$-parts.

The lateral flux
in right part of Eq.\,(\ref{difcons}) is due to $(z,r)$-component of the
energy-momentum $
{\sf f}^{z}\equiv \ds \frac{d}{dt}\int { T}^{zr} dS_{r}$ and consists of antisymmetric and branon part:
\begin{align}\label{lfl}
&{\sf f}^{z}_{\ah}  = -k m\gamma\,\sgn(t)\left[(D-3){\kern
1pt}v^2+1\right],\qquad
    {
    \sf
f}^{z}_{\bh}  =- \delta \dot{P}^z_{\bh} \,.
\end{align}
 All the momenta derivatives are
constant before and after the  moment of piercing \mbox{$t=0$} when they
change the sign.  The sum \mbox{$\delta \dot P^M+\delta \dot{\bar{P}}^M$} does not vanish
for both values of $M$. This is not surprising since we still need
to add contribution of the gravitational stresses.

\medskip

{\bf Gravitational dressing.}  Analysing different terms in
$S^{MN}(h,\bar h)$ obtained by substituting the metric deviations
 $h_{MN}$  and  $\bar{h}_{MN}$ one detects  presence of
contributions of two types: containing the delta-function
 $\delta(z)$  localized in the wall world-volume $ S $, and
containing the delta-function $\delta^D\!\left(x- u \tau\right)$
localized at the particle world-line  $ \bar S $. The corresponding
momentum time derivatives  for the particle are:
\begin{align}\label{stressp}
& \bar{f}^z\equiv \dot{\bar S}^z = -2 \left(D+1\right)m k \gamma v^2 \,\sgn
(t)\,,\non\\
& \bar{f}^0\equiv \dot{\bar{S}}^0= -2 \left(\vp  \left(D+1\right) v^2
+\fr{4}{\ga^2}\right)
    m k\ga v \,\sgn (t)\,,
 \end{align}
while the corresponding brane components read:
\begin{align}\label{stressd}
& {f}^z \equiv \delta \dot{P}^z_{S} = \ga v^2
 \left[(D-2)\ga^2v^2+3\vp \right] m k \,\sgn(t)
    \,,\non\\&
   {f}^0\equiv   \dot{ {S}}^0=\ga v \left[(D-2)\ga^2v^2+2D-5- \frac{2(D-3)}{\ga^{2}}\right] \,\sgn (t)\,.
 \end{align}

  The dressed particle energy term
is \mbox{$ \ds\dot{\bar{{\cal P}}}^0
={\delta\dot{\bar{P}}^{0}}+\bar{f}^0 $}, while the dressed brane
energy term is  \mbox{$\ds
 {\dot{\cal {P}}}^0 =\delta\dot{ {P}}^{0}+{f}^0  $}.
 It is easy to establish
 the conservation equation
 \mbox{$\ds {\dot{\bar{\cal {P}}}}^0+ \dot{\cal P}^0 =0$} . Note that
\mbox{$ \delta\dot{ {P}}^{0}+\delta\dot{\bar{P}}^{0}\neq 0$}. Thus
 treating gravity perturbatively in Minkowski space, one is able to
introduce, instead of the  potential  energy, the  dressed
quantities for the particle in the gravitational field of the domain
wall, and respectively, of the domain wall in the gravitational
field of the particle  such that their sum is conserved.

The situation of the spatial $z$- component is more involved. The
dressed brane $z$-momentum terms can be defined analogously as
\mbox{$\ds\dot{\cal{P}}^z= \delta\dot{ {P}}^{z} +f^z $} for the
brane and
 \mbox{$\ds {\dot{\bar{\cal{P}}}}^z= \delta\dot{\bar{P}}^{z}+{\bar{f}}^z$}.
for the particle. Their sum, however, is not zero. The balance
equation contains the lateral flux:
   $$ \dot{\cal{P}}^z+{\dot{\bar{\cal{P}}}}^z=-f^z\,. $$
Note that, since $\sgn(t)$ and $ \theta(t)$ are linearly
independent, this extra contribution holds separately for $a$ and
$b$ branon components.
 \section{Concluding remarks}
We have shown that non-local gravitational stresses effectively
localize within perturbative treatment of gravity enabling us to
establish the local energy-momentum conservation for the scattering
problem without asymptotical free states.  The unusual  feature of
the balance equation is the existence of non-zero flux of
$z-$component of the momentum density through the lateral surface of
the world-tube  due to the branon. Strictly speaking, in this case
the conserved momentum  can not be defined as an integral over the
space only. But if one still keeps such a definition, the imbalance
will be exactly accounted for by the lateral flux  playing the role
of an external force. This takes place separately for the bound part
of the branon, and for the free branon wave.

\bigskip
\noindent {\bf Acknowledgments.} This work was supported by the RFBR
grant 14-02-01092a.  PS also acknowledges support   by the EU
programs ``Thales" (MIS 375734) and  ``ARISTEIA II", and by the
non-commercial ``Dynasty'' Foundation (Russian Federation).


\begin{thebibliography}{10}

\bibitem{linear}
 A.~Vilenkin,
 Phys.\ Rev.\ D {\bf 23}, 852 (1981);
 Phys.\ Rep.\ {\bf 121}, 263 (1985).

\bibitem{Vilsh}
 A.~Vilenkin and E.~P.~S.~Shellard,
 {\sl Cosmic Strings and other Topological Defects},
 Cambridge University Press, Cambridge (2000).

\bibitem{ChamEard}
 A.~Chamblin and D.~M.~Eardley,
  Phys.\ Lett.\  B {\bf 475}, 46 (2000).

\bibitem{Stojkovic:2005zh}
  D.~Stojkovic, K.~Freese and G.~D.~Starkman,
  Phys.\ Rev.\ D {\bf 72}, 045012 (2005).

\bibitem{Flachi:2007ev}
  A.~Flachi and T.~Tanaka,
  Phys.\ Rev.\ D {\bf 76}, 025007 (2007).

\bibitem{GaMeS1}
D.\,V.\,Gal'tsov, E.\,Yu.\,Melkumova, P.\,Spirin,
  Phys.\,Rev.\,D \textbf{89}, 085017
(2014).

\bibitem{GaMeS2}
D.\,V.\,Gal'tsov, E.\,Yu.\,Melkumova, P.\,Spirin,
Phys.\,Rev.\,D \textbf{90}, 125024 (2014).

\end{thebibliography}
\end{document}